\documentclass[a4paper,11pt]{article}
\pdfoutput=1 

\usepackage{jheppub} 

\usepackage[usenames,dvipsnames]{xcolor}
\usepackage{hyperref}
\hypersetup{
    pdfencoding=unicode,
	colorlinks=true,
	urlcolor=Maroon,
	linkcolor=RoyalBlue,
	citecolor=OrangeRed,
	pdfdisplaydoctitle=true,
	pdfstartview=FitH,
	linktocpage=true
}
\usepackage{mathtools}
\usepackage[skins]{tcolorbox}

\title{\boldmath Remarks on the critical dimension of the left-handed string and its quasiconformal nature}

\author[a]{Yuqi Li}
\author[a]{and Warren Siegel}

\affiliation[a]{C.~N.~Yang Institute for Theoretical Physics,\\ State University of New York, Stony Brook, NY 11794-3840}

\emailAdd{yuqi.li@stonybrook.edu}

\preprint{YITP-SB-2022-18}

\abstract{The chiral string without a singular gauge limit is argued to have the same critical dimension as its corresponding conventional closed string. Thus, its central charge would be the same as its conventional counterpart in the conformal gauge. Here, we would re-examine the critical dimension of the chiral string in the singular Hohm-Siegel-Zwiebach limit. A straight forward calculation of the operator product expansion (OPE) of the corresponding would-be stress tensor shows that the central charge term is not the same as its conventional counterpart when taking the singular limit. Instead of having a conformal transformation on the worldsheet, the coordinate reparametrization provides a set of quasiconformal mappings.}

\begin{document} 
\maketitle
\flushbottom

\section{Introduction}
In \cite{Siegel:2015axg}, Siegel shows that the Cachazo-He-Yuan (CHY) formula is reproduced from string theory by both changing the boundary condition and taking the singular Hohm-Siegel-Zwiebach (HSZ) gauge limit. Without the singular gauge limit \cite{Huang:2016bdd}, one can reproduce the massless truncation of ordinary string theory with only the change of boundary condition as well. From now on, we use the term ``left-handed string" to refer to the string with both changed boundary conditions and singular limit, whereas the ``chiral string" stands for the string with only the change of boundary conditions. Further calculations (\cite{Leite:2016fno, Lee:2017utr, Casali:2017mss,Azevedo:2019zbn,Guillen:2021mwp,Guillen:2021nky,Jusinskas:2021bdj}) also show that it is almost obvious that the corresponding critical dimension of chiral string is the same as its conventional counterpart in string theory. 

One might think that it is straight forward to restore the similar statement on the critical dimension of left-handed string. However, it is less obvious than it seems to be. On the one hand, the left-handed string can reproduce the complete field theory result by means of the CHY formula \cite{Li:2017emw}, which is valid for arbitrary spacetime dimension; on the other hand, the existence of critical dimension of the left-handed string will ruin the validity of the formula at non-critical spacetime dimension. Thus, it is important to check whether the critical dimension of left-handed string is the same as chiral string. 

To examine this, we compute the corresponding would-be stress tensor of the left-handed string and calculate the operator product expansion (OPE) of those stress tensors. After taking the singular limit, the would-be central charge term contributes trivially, and the rest terms behave chirally since the left-handed string has only one handedness. Therefore, it is sufficient to say that the left-handed string, effectively, has no critical dimension and provides a consistent field theory limit, which paves the way for using the left-handed string to calculate the infrared singularities of scattering amplitudes at higher genus.

It is also fascinating that the coordinate reparametrization without the singular limit gives a set of one parameter valued quasiconformal mappings. When taking the singular limit of the parameter, one almost 1-quasiconformal (conformal) mapping shows up again.

\section{Gauge choices}
As is shown in \cite{Siegel:2015axg}, two ingredients are needed for the recipe of left-handed string:
\begin{itemize}
    \item Change of boundary conditions of conformal fields on the worldsheet, which effectively gives rise to a relative minus sign change in Green's function.\footnote{Based on this set up, one can also call it twisted string or sectorized string \cite{Casali:2017mss,Azevedo:2019zbn,Guillen:2021mwp,Guillen:2021nky,Jusinskas:2021bdj}}
    \item Reparametrize the worldsheet coordinates with a gauge parameter and take the singular limit of the gauge parameter to obtain the HSZ limit, which effectively makes the whole theory left-handed.
\end{itemize}
Let us start from the bosonic Lagrangian in Hamiltonian form of string on the worldsheet, with some Lagrangian multipliers and setting $\alpha^\prime=1$,
\begin{eqnarray}
\mathcal{L}_H=-\dot{X}\cdot P+\frac{\sqrt{-h}}{h_{11}}\Big[\frac{1}{2}(P\cdot P+X^\prime\cdot X^\prime)\Big]+\frac{h_{01}}{h_{11}}\Big[X'\cdot P\Big]
\end{eqnarray}
here, $A\cdot B=\eta^{\mu\nu}A_\mu B_\nu$ with flat metric $\eta^{\mu\nu}$ and $h$ is the determinant of the worldsheet metric with fixed diffeomorphisms. We now set
$$\lambda_0=\frac{\sqrt{-h}}{h_{11}}\,\,\,\, \text{and}\,\,\,\, \lambda_1=\frac{h_{01}}{h_{11}}$$
as the same as in \cite{Siegel:2015axg}. Thus the corresponding Lagrangian could be written as:
\begin{eqnarray}\label{eq:action}
\mathcal{L}_L=-\frac{1}{2}\Big[\partial_LX(z_L,z_R)\Big]\cdot\Big[\partial_RX(z_L,z_R)\Big]
\end{eqnarray}
with 
\begin{eqnarray}\label{eq:gauge}
\partial_L&=&\frac{1}{\sqrt{\lambda_0}}[\partial_\tau-(\lambda_1-\lambda_0)\partial_\sigma]=\frac{1}{\sqrt{1+\beta}}(\partial+\beta\bar{\partial})\nonumber\\
\partial_R&=&\frac{1}{\sqrt{\lambda_0}}[\partial_\tau-(\lambda_1+\lambda_0)\partial_\sigma]=\sqrt{1+\beta}\bar{\partial}
\end{eqnarray}
Thus the worldsheet coordinates gain some reparametrizations,
\begin{eqnarray}\label{eq:coordinate}
z_L&=&\frac{1}{2}\frac{1}{\sqrt{\lambda_0}}[(\lambda_1+\lambda_0)\tau+\sigma]=\sqrt{1+\beta}z\nonumber\\
z_R&=&-\frac{1}{2}\frac{1}{\sqrt{\lambda_0}}[(\lambda_1-\lambda_0)\tau+\sigma]=\frac{1}{\sqrt{1+\beta}}(\bar z-\beta z)
\end{eqnarray}
here, $\beta$ is the gauge parameter,
\begin{eqnarray}
\lambda_0=\frac{1}{1+\beta},\,\,\,\lambda_1=\frac{\beta}{1+\beta};\,\,\,\beta\ge0
\end{eqnarray}
When taking the singular limit $\beta\rightarrow\infty$, the HSZ theory restores as we mentioned above, and $\beta=0$ gives back the conventional string with the full conformal symmetry recovered. With the help of (\ref{eq:gauge}), the Lagrangian (\ref{eq:action}) can be re-written into a simpler form:
\begin{eqnarray}\label{eq:lag}
\mathcal{L}_L=-\frac{1}{2}\Big[\beta(\bar\partial X)\cdot(\bar\partial X)+(\partial X)\cdot(\bar\partial X)\Big]
\end{eqnarray}
It is not so hard to see that one handedness manifests in the Lagrangian when taking the singular $\beta\rightarrow\infty$ limit.

\section{Would-be stress tensor and OPE}
Noticed that the reparametrization (\ref{eq:coordinate}) is not changing the form of the worldsheet action (\ref{eq:action}). Without taking the singular limit $\beta\rightarrow\infty$, there is still a would-be stress tensor corresponding to the worldsheet action. We name those stress tensors as $T_L$ and $T_R$ with respect to $T_{zz}$ and $T_{\bar{z}\bar{z}}$ of the original worldsheet conformal field theory; namely, the OPE of the reparametrized stress tensor can be formally written into
\begin{eqnarray}
T_L(z_L)T_L(w_L)&\sim&\frac{c_L/2}{(z_L-w_L)^4}+\frac{2T_L(w)}{(z_L-w_L)^2}+\frac{\partial_L T_L(w)}{(z_L-w_L)}\label{eq:OPE_L}\\
T_R(z_R)T_R(w_R)&\sim&\frac{c_R/2}{(z_R-W_R)^4}+\frac{2T_R(w)}{(z_R-w_R)^2}+\frac{\partial_RT_R(w)}{(z_R-w_R)}\label{eq:OPE_R}
\end{eqnarray}
As mentioned before, the would-be central charges $c_{L}$ and $c_R$ corresponding to the would-be stress tensor should be the same as the corresponding original string; that is, $c_L=c_R=c$, with $c$ the central charge of any conventional closed string before reparametrization. Noted that the reparametrization is implicit here. 

Now treat reparametrization (\ref{eq:coordinate}) as the pushforward from the local coordinate $(z,\bar{z})$ to local coordinate $(z_L,z_R)$. In the new local coordinate, the induced stress tensor is the would-be stress tensor:
    \begin{eqnarray*}
    T_L(z_L)=\partial_LX(z_L)\cdot\partial_LX(z_L)\coloneqq[\partial_LX\cdot\partial_LX](z_L)\\
    T_R(z_R)=\partial_LX(z_R)\cdot\partial_LX(z_R)\coloneqq[\partial_RX\cdot\partial_RX](z_R)
    \end{eqnarray*}
Use the coordinate reparametrization (\ref{eq:coordinate}) and the would-be stress tensors transform as
    \begin{eqnarray}
    T_L(z_L)&=&\frac{1}{1+\beta}[\partial X\cdot\partial X+2\beta\partial X\cdot\bar\partial X+\beta^2\bar\partial X\cdot\bar\partial X](z_L)\label{eq:str_L}\\
    T_R(z_R)&=&(1+\beta)[\bar\partial X\cdot\bar\partial X](z_R)\label{eq:str_R}
    \end{eqnarray}
With the help of the coordinates reparametrization (\ref{eq:coordinate}), one get
\begin{eqnarray*}
T_L(z_L)T_L(w_L)&\sim&\frac{c/2}{(1+\beta)^2(z-w)^4}+\frac{2T_L(w)}{(1+\beta)(z-w)^2}+\frac{\partial_L T_L(w)}{\sqrt{1+\beta}(z-w)}\\
T_R(z_R)T_R(w_R)&\sim&\frac{(1+\beta)^2c/2}{[(\overline{z-w})-\beta(z-w)]^4}+\frac{2(1+\beta)T_R(w)}{[(\overline{z-w})-\beta(z-w)]^2}+\frac{\sqrt{1+\beta}\partial_RT_R(w)}{(\overline{z-w})-\beta(z-w)}
\end{eqnarray*}
Plug (\ref{eq:str_L}) and (\ref{eq:str_R}) in the right-hand side and expand to the leading order in the large $\beta$ limit. After some straight forward calculations, one can easily get\footnote{Noted that the inverse mapping of (\ref{eq:gauge}) has been used in the calculations of the last term of (\ref{eq:tr_L}) and (\ref{eq:tr_R}).}
\begin{eqnarray}
T_L(z)T_L(w)&\sim&\lim_{\beta\gg1}\Big[\frac{c/2}{(1+\beta)^2(z-w)^4}+\frac{\beta^2}{(1+\beta)^2}\frac{2\bar T(\bar z)}{(z-w)^2}-\frac{\beta^2}{(1+\beta)\beta}\frac{\bar\partial \bar T(\bar z)}{z-w}\Big]\\\label{eq:tr_L}
T_R(z)T_R(w)&\sim&\lim_{\beta\gg1}\Big[\frac{(1+\beta)^2}{\beta^4}\frac{c/2}{(z-w)^4}+\frac{(1+\beta)^2}{\beta^2}\frac{2\bar T(\bar z)}{(z-w)^2}-\frac{(1+\beta)}{\beta}\frac{\bar\partial \bar T(\bar z)}{z-w}\Big]\label{eq:tr_R}
\end{eqnarray}
here, define $T(z)\coloneqq\partial X\cdot\partial X$ and $\bar T(z)\coloneqq\bar\partial X\cdot\bar\partial X$ as usual. When taking central charge finite as the same as its corresponding conventional closed string theory, the first term of both (\ref{eq:tr_L}) and (\ref{eq:tr_R}) behave trivially while $\beta\rightarrow\infty$.
That is,
\begin{eqnarray}
\lim_{\beta\rightarrow\infty}T_L(z)T_L(w)&\sim&\frac{2\bar T(\bar z)}{(z-w)^2}-\frac{\bar\partial \bar T(\bar z)}{z-w}\\
\lim_{\beta\rightarrow\infty}T_R(z)T_R(w)&\sim&\frac{2\bar T(\bar z)}{(z-w)^2}-\frac{\bar\partial \bar T(\bar z)}{z-w}
\end{eqnarray}
Then, use the fact of one handedness of left-handed string such that $\frac{1}{z-w}\sim-\frac{1}{\bar z-\bar w}$ when $z\rightarrow w$ at $\beta\rightarrow\infty$ limit. The OPE becomes completely chiral:
\begin{eqnarray}
\lim_{\beta\rightarrow\infty}T_L(z)T_L(w)&\sim&\frac{2\bar T(\bar z)}{(\bar{z}-\bar{w})^2}+\frac{\bar\partial \bar T(\bar z)}{\bar z-\bar w}\\\label{eq:OPE_Lf}
\lim_{\beta\rightarrow\infty}T_R(z)T_R(w)&\sim&\frac{2\bar T(\bar z)}{(\bar z-\bar w)^2}+\frac{\bar\partial \bar T(\bar z)}{\bar z-\bar w}\label{eq:OPE_Rf}
\end{eqnarray}
As a consistent check, one can get similar results directly from some careful calculations by taking the singular limit of the Lagrangian (\ref{eq:lag}) as well.
\section{From conformal to quasiconformal}
Recall the action (\ref{eq:action}) in the local coordinate $(z_L,z_R)$, one can write the field into holomophic functions with respect to $z_L$ and $z_R$,
$$X(z_L,z_R)=X_L(z_L)+X_R(z_R)$$
The holomorphicity condition requires,
\begin{eqnarray}
\partial_L X_R(z_R)=0\,\,\,\,\,\text{and}\,\,\,\,\,\,\partial_RX_L(z_L)=0
\end{eqnarray}
Now the reparametrization (\ref{eq:coordinate}) and (\ref{eq:gauge}) gives us,
\begin{eqnarray}\label{eq:quasi}
\frac{\partial X_R}{\partial \bar z}=-\frac{1}{\beta}\frac{\partial X_R}{\partial z}
\end{eqnarray}
Then set $\mu(z)=-\frac{1}{\beta}$. Noted that $|\mu(z)|<1$ valid when $\beta\gg 1$ and define
$$K(z)=\frac{1+|\mu(z)|}{1-|\mu(z)|}=\frac{1+\frac{1}{\beta}}{1-\frac{1}{\beta}}$$
Thus, the field $X_R(z_R)$ is $K$-quasiconformal in the local $(z,\bar z)$ coordinate. At the singular $\beta\rightarrow\infty$ limit,
\begin{eqnarray}
\lim_{\beta\rightarrow\infty}K(z)=\frac{\beta+1}{\beta-1}\rightarrow1
\end{eqnarray}
Namely, in the singular $\beta\rightarrow\infty$ limit, the theory is reduced to 1-quasiconformal, which is conformal. However, the resulting theory is not the same as the original one since it has no central charge term as discussed in the previous section.

\section{Conclusion and discussion}
As a summary, the would-be stress tensor OPE of left-handed string does not have a non-trivial central charge term; that is, no critical dimension is effectively needed for the calculations of the left-handed string.
Indeed, as mentioned in \cite{Siegel:2015axg}, the would-be stress tensor of ghosts fields, $T^{gh}$, receives corrections from the coordinate reparametrization (\ref{eq:coordinate}) and (\ref{eq:gauge}) and one extra $\sqrt{1+\beta}$ factor because of the OPE $b(z)c(0)\sim \frac{1}{z}$, namely $T^{gh}_L(z)\rightarrow\sqrt{1+\beta}T^{gh}(z_L)$. Then the OPE gives 
\begin{eqnarray*}
T^{gh}_L(z)T^{gh}_L(0)&\sim&\lim_{\beta\rightarrow\infty} \Big[\frac{c_L^{gh}/2}{(1+\beta)^2z^2}+\frac{\beta2\bar T^{gh}(z)}{(1+\beta)\bar z^2}+\frac{\beta\bar\partial \bar T^{gh}(z)}{(1+\beta)\bar z}\Big]\\
&\sim&\frac{2\bar T^{gh}(z)}{\bar z^2}+\frac{\bar\partial \bar T^{gh}(z)}{\bar z}
\end{eqnarray*}
and also
\begin{equation*}
    T^{gh}_R(z)T^{gh}_R(0)\sim\frac{2\bar T^{gh}(z)}{\bar z^2}+\frac{\bar\partial \bar T^{gh}(z)}{\bar z}.
\end{equation*}
Thus the would-be ghost field central charge terms will not contribute non-trivially. Meanwhile, when $\beta$ is not reaching the infinity, the sum of would-be central charge terms of both matter fields and ghosts with the same chirality can be written into
$\frac{c_L+c_L^{gh}}{(z_L-w_L)^4}$ and $\frac{c_R+c_R^{gh}}{(z_R-w_R)^4}$
such that the critical dimension does not depend on $\beta$, which agrees with \cite{Leite:2016fno, Lee:2017utr, Casali:2017mss,Azevedo:2019zbn,Guillen:2021mwp,Guillen:2021nky,Jusinskas:2021bdj}.\footnote{One can also do the similar calculation to show that the would-be central charge terms cannot contribute non-trivially in supersymmetric theories as well.}
Without the obstruction of the existence of critical dimension, the scattering amplitude of left-handed string is effectively valid in arbitrary dimension; one can do the $\bar z$-integral such that the remaining conformal fields of left-handed string (correpongding to $\bar z$) are completely decoupled. The left-over $\bar z$-independent part would reproduce a field theory result, namely CHY formula. A graph below shows how it works,
\begin{eqnarray*}
\{\text{conf}\oplus\overline{\text{conf}}\}\xRightarrow{\text{$\beta\neq0$}}\{\text{conf}\oplus\overline{\text{quasiconf}}\}\xRightarrow{\beta\rightarrow\infty}\{(\text{field}\otimes\overline{\text{conf}})\oplus\overline{\text{conf}}\}\xRightarrow{\int \text{d}\bar z}\text{CHY}
\end{eqnarray*}
here, ``$\text{conf}$" and ``$\text{quasiconf}$" stand for ``conformal" and ``quasiconformal", respectively, and the barred and un-barred are for different chiralities. Thus, it is sufficient to say that
\begin{tcolorbox}[enhanced,width=5in,center upper,
    fontupper=\large\bfseries,drop shadow southwest,sharp corners]
CHY$=$Left-handed string amplitudes
\end{tcolorbox}
Noted that the calculation based on quasiconformal mapping can only show that a conformal field theory is reproduced while $\beta\rightarrow\infty$. So, one has to use the coordinate reparametrization to see whether the reproduced conformal field theory has a non-trivial central charge term. In \cite{Siegel:2015axg}, instead of our (\ref{eq:coordinate}), Siegel used $z_L\rightarrow z\,\,\,\,\text{and} \,\,\,\,\,z_R\rightarrow\bar z-\beta z.$
It would not change the result in string amplitude calculations since the coordinates always come together pairwise. However, the extra $\sqrt{1+\beta}$ factor matters in our calculations as the coordinates come up chirally in OPEs.

In the future work, it would be very interesting to see the relationships between left-handed string amplitudes and tropical amplitudes in higher genus \cite{Tourkine:2013rda,Arkani-Hamed:2022cqe}.

\acknowledgments
Our work is supported by NSF grant PHY-1915093. We thank Martin Ro\v cek for useful discussions and suggestions. YL would also like to deliver special thanks to Xiaoying Xie for her review of the very first draft of this paper.

\end{document}